\documentclass{ifacconf}

\usepackage{graphicx}      
\usepackage{natbib}        
\usepackage{caption}
\usepackage{subcaption}
\usepackage{floatrow}
\usepackage{graphicx}
\usepackage{amsmath}
\usepackage{amssymb}
\usepackage{multirow}
\usepackage{multicol}%
\usepackage{soul}

\newcommand{\SEI}{\mathrm{SEI}}
\newcommand\R{\mathrm}
\begin{document}
\begin{frontmatter}

\title{ The Case for DeepSOH: 
Addressing Path Dependency for Remaining Useful Life }

\author[First]{Hamidreza Movahedi} 
\author[First]{Andrew Weng}
\author[First]{Sravan Pannala}
\author[First]{Jason B. Siegel}
\author[First]{Anna G. Stefanopoulou} 

\address[First]{Battery Controls Group, University of Michigan, Ann Arbor, MI, 48109 USA (e-mail:  movahedi@umich.edu).}

\begin{abstract}                

The battery state of health (SOH) based on capacity fade and resistance increase is not sufficient for predicting Remaining Useful life (RUL). The electrochemical community blames the path-dependency of the battery degradation mechanisms for our inability to forecast the degradation. 

The control community knows that 
the path dependency is addressed by 
full state estimation. 

We show that even the electrode-specific SOH (eSOH) estimation is not enough to fully define the degradation states by simulating infinite possible degradation trajectories and remaining useful lives (RUL) from a unique eSOH. 
We finally define the deepSOH states that capture the individual contributions of all the common degradation mechanisms, namely, SEI, plating, and mechanical fracture to the loss of lithium inventory. 
We show that the addition of cell expansion measurement may allow us to estimate the deepSOH and predict the remaining useful life. 
\end{abstract}


\end{frontmatter}

\section{Introduction}

The ability to accurately estimate the battery state of health (SOH) and remaining useful life (RUL) remains a critical yet underdeveloped feature of modern battery management systems (BMS). The RUL is especially important for understanding the feasibility of second-life battery applications, including for grid storage and electric vehicle resale. Unfortunately, accurate RUL prediction remains incredibly challenging. The inability to predict RUL of aged vehicles has already begun to impact the marketplace. A recent study showed that the average price of an EV fell 31.8\% in the past year compared to just 3.6\% for internal combustion engine vehicles \citep{Butts2024-ed}. The poor resale value is partly owed to the lack of long-term reliability data, i.e. RUL, of these electric vehicles.

Recognizing the need for improved RUL estimation, the research community has been broadening the definition of SOH to include metrics beyond cell-level capacity fade \citep{mohtat2017identifying}. Traditionally, the SOH is defined simply as
\begin{equation}
    \mathrm{SOH} = C/C_\mathrm{nom}
\end{equation}
where $C$ is the cell capacity measured at the time of testing and $C_\mathrm{nom}$ is the initial capacity. The same concept has been used for defining SOH based on resistance or power level.
\cite{mohtat2019towards} proposed an extended definition of SOH, one which involves electrode-specific measures, or `eSOH', according to:
\begin{equation}
    \mathrm{eSOH} \triangleq [C, C_p, C_n, y_0,x_0]^T
\end{equation}
where $C_n$ and $C_p$ are the negative and positive electrode capacities, and $x_0$ and $y_{0}$ are the lithium stochiometries at 0\% SOC. From these state variables, electrode-specific (i.e. ``component-level'') degradation modes can be calculated directly from these state variables  according to
\begin{align}
    \mathrm{LAM}^+ &= 1 - C_p/C_{p,\mathrm{nom}} \\   
    \mathrm{LAM}^- &= 1 - C_n/C_{n,\mathrm{nom}} \\
    \mathrm{LLI} &= 1 - n_\mathrm{Li}/n_{\mathrm{Li},\mathrm{nom}}
\end{align}
where LLI indicates the loss of lithium inventory, LAM indicates the loss of active material, and the variables denoted by `nom' indicate nominal values, and the remaining variables are obtained at the time of measurement \citep{ lee2020electrode}. A major strength of the eSOH approach is thus that it enables component-level understanding of the origin of cell degradation. The eSOH definition continues to see adoption for state estimation applications, including recently by \cite{lopetegi2024new}.

However, despite the relative success of the eSOH approach, a major limitation of the eSOH approach is that it cannot distinguish between different degradation mechanisms contributing to LLI. In particular, LLI can arise from different sources. For example, if the growth of the solid-electrolyte interphase (SEI), lithium plating, and a source of active material loss (such as mechanical degradation from particle fracture) are present in a cell:
\begin{equation}
    {LLI} = {LLI}_\mathrm{SEI} + {LLI}_\mathrm{pl} + {LLI}_\mathrm{LAM}.    
\end{equation}

While all LLI mechanisms result in capacity loss, an accurate RUL estimator must be able to identify the LLI components separately. For example, lithium plating is a more significant contributor to RUL since lithium plating is believed to be a primary cause of ``knees'' in degradation trajectories \citep{Attia2022-mf}, therefore determining the LLI due to plating ($LLI_\mathrm{pl}$) is necessary for predicting the RUL.

This work advocates for extending the definition of SOH to include states of the degradation mechanisms, such as thickness expansion due to SEI ($\delta_\mathrm{SEI}$) and thickness expansion due to lithium plating ($\delta_\mathrm{pl}$). We call the augmented system ``deep SOH,'' defined by
\begin{equation}
    \mathrm{deepSOH} \triangleq [\delta_{SEI}, \delta_{pl}, C_{p}, C_{n}, LLI]^T
    \label{deepSOH}
\end{equation}

\begin{figure*}[h]
\begin{center}
\includegraphics[width=13cm]{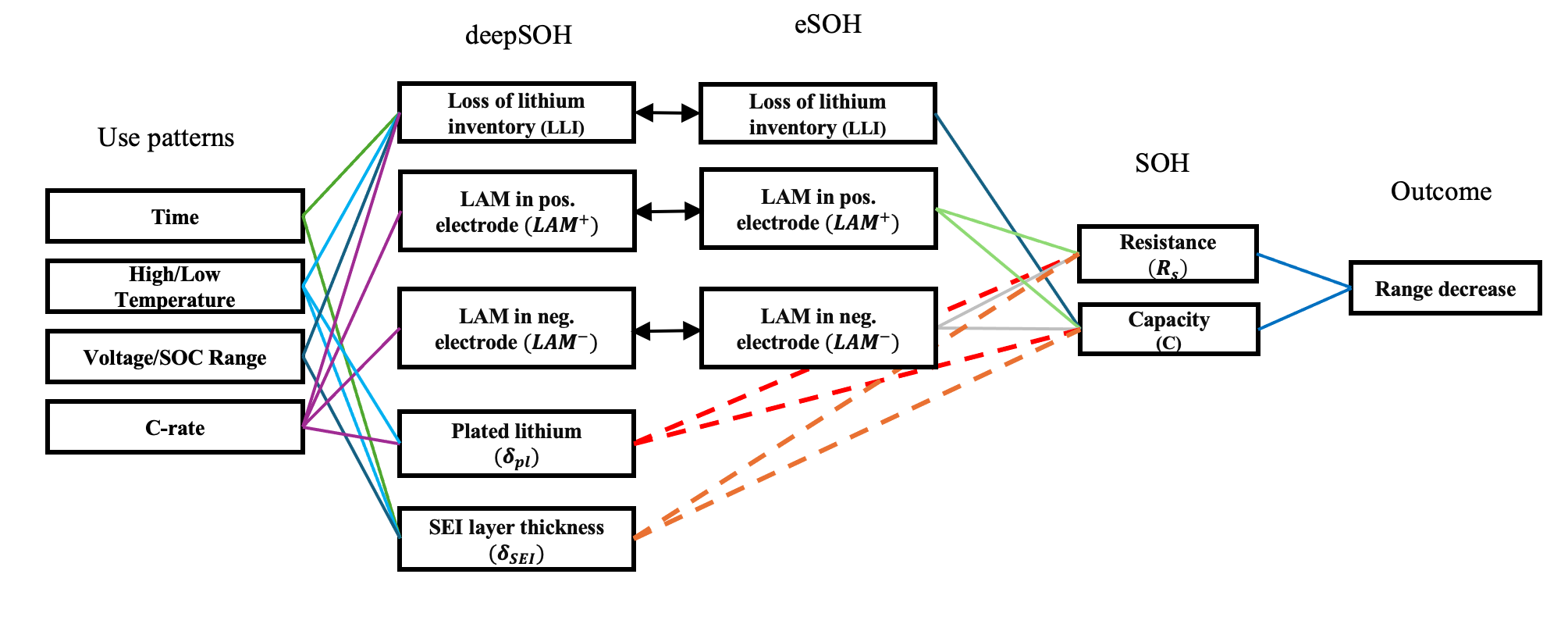}    
\caption{ Schematic of different SOH definitions and their relationship with the use patterns and outcome. The dashed lines are ignored by eSOH.}
\label{fig: deepSOH blocks}
\end{center}
\end{figure*}

Figure \ref{fig: deepSOH blocks} highlights that the two additional states represented by deepSOH are critically linked to the cell capacity and resistance, and ultimately determine vehicle range for electric vehicle applications. A description of $\delta_{pl}$ and $\delta_\mathrm{SEI}$ is necessary to provide a more complete picture of RUL. In this work, we show that while efforts to estimate the eSOH \citep{lopetegi2024new,lee2020electrode} are necessary they are not sufficient to provide a full picture of the path dependency of the cell's degradation and to predict RUL accurately. Specifically, we show that cells having the same eSOH and resistance states but different deepSOH can have different RULs. We further show that identifying the deepSOH (for instance, by measuring the irreversible expansion of the cells) improves the identifiability of the degradation mechanisms.

We assert that a deepSOH is necessary to resolve the mechanistic origins of LLI and improve RUL estimation. We show, through analytical calculations and numerical simulations,
that identifying these states improves RUL estimation accuracy.

The paper is structured as follows. Section 2 provides a brief description of the degradation model. Section 3 further clarifies the degradation dynamic system and introduces the concept of deepSOH. Section 4 clarifies the necessity of deepSOH through analytical derivation and simulation results. Section 5 finally concludes the paper by highlighting the takeaways, limitations, and next steps.
\section{Battery degradation model}
The degradation models we consider here are based on a single-particle model (SPM), where the electrolyte dynamics are ignored and each electrode is modeled with a spherical particle. The degradation mechanisms considered here are SEI layer growth, mechanical degradation, and lithium plating. We have shown the full description of the degradation models in  \citep{pannala_movahedi_garrick_siegel_stefanopoulou_2023} where we were able to model capacity degradation, instantaneous resistance, and irreversible expansion at the same time. Here we briefly describe each mechanism with the aim of identifying the states in each one.
\subsection{SEI layer growth}
Growth of the solid-electrolyte interphase at the negative electrode-electrolyte interface is ubiquitous in modern lithium-ion batteries and is responsible for continuous capacity fade. 
These SEI growth can be modeled \citep{movahedi2023physics} by:
\begin{equation}
    \label{SEI system eqs}
    \left\{ 
    \begin{array}{l}
        j_{SEI} = -k_{SEI}  c_{EC}^s \exp\left(-\frac{\alpha_\R{SEI}F}{RT}\eta_\R{SEI}\right) \\
        \eta_{SEI}=\eta^-+U^-\left(c_{ss}^-\right) -U_{SEI}\\
        \frac{\partial c_\SEI}{\partial t} = -\frac{a^-_s j_{SEI}}{2}, \ \delta_{SEI}=\frac{C_{SEI} \Omega_{SEI}}{a_s^-}\\
        -D_\R{SEI} \frac{c_{EC}^s-c_{EC}^0}{\delta_{\SEI}} = -j_\R{SEI}
    \end{array} 
\right.
\end{equation}
where $c_\SEI$ is the concentration of formed SEI, $D_{\SEI}$ is the diffusivity of SEI layer, $k_\SEI$ is the kinetic rate constant, $\eta^-$ is the negative electrode overpotential corresponding to the intercalation reaction, $U^-\left(c_\R{ss}^-\right)$ is the surface concentration-dependent negative electrode open circuit potential (OCP), and $U_\R{SEI}$ is constant the potential of SEI reaction.

 The specific surface area of active material in the negative electrode is calculated  $a_s^- =3\varepsilon_{s}/r_{p}^-$. Where $r_{p}^-$ is electrode particle radius and the volume fraction of active material is calculated as $
    \varepsilon_{s}^\pm = \frac{3600 \: C_{p/n}}{AFl^\pm c_{s,max}^\pm}$.

In this equation, $A$ is the cell surface area, $l^\pm$ is the thickness of the electrode, and $c_{s, max}^\pm$ signifies the maximum concentration in each electrode.

Streamlining the equations in \ref{SEI system eqs}, the overall reaction rate is the combination of a kinetic and diffusion-limited mechanism that governs the SEI growth:
\begin{flalign}
     j_\R{SEI} = \frac{-c_{EC}^0}{1/(k_\R{SEI}\exp\left(\frac{-\alpha_\R{SEI}F}{RT}\eta_\R{SEI}\right)) + \delta_{\SEI}/(D_\R{SEI})}.
     \label{jsei}
\end{flalign}
The SEI layer growth rate can be presented as:
\begin{flalign}
    \dot{\delta}_{SEI}= - \frac{\Omega_{SEI} \: j_{SEI}}{2}
    \label{del dot sei}
\end{flalign}
We can simplify the notations of Eqs.(\ref{jsei}, \ref{del dot sei}) by writing:
\begin{flalign}
    \dot{\delta}_{SEI}= f_{1}(\delta_{SEI}, C_n,LLI,I)
    \label{del dot sei f1}
\end{flalign}
where $I$ is the current input and affects the $\eta_{SEI}$. The function $f_1(.,.,.,.)$ also includes LLI, as the cyclable lithium affects the surface concentration $c_{ss}$.
The moles of lithium consumed during the SEI reaction can be computed as:
\begin{flalign}
    \Delta n_{Li, SEI}= -\frac{2 A l^- a_{s}^- }{\Omega_{SEI}}\delta_{SEI}
    \label{moles SEI}
\end{flalign}

 \subsection{Lithium plating}

 Lithium plating occurs when the negative electrode is fully occupied by lithium or high-charge pulses. Lithium plating is a critical degradation mechanism to model since it can lead to internal shorts, battery failure, and thermal runaway. 
 The concentration of total plated lithium and the corresponding film thickness can also be modeled simply according to the following set of equations:
\begin{equation}
    \label{plating system eqs}
    \left\{ 
    \begin{array}{l}
        j_{pl} = -\frac{k_{pl} c_e (c_{ss}^--c_{avg}^-)}{c_{s,amx}^-} \exp\left(-\frac{\alpha_\R{pl}F}{RT}\eta_\R{pl}\right) \\
       \eta_\R{pl}=\eta^-+U^-\left(c_\R{ss}^-\right)\\
        \frac{\partial c_{pl}}{\partial t} = a^-_s j_{pl}, \ \delta_{pl}=\frac{c_{pl} \Omega_{pl}}{a_s^-}
    \end{array} 
\right.
\end{equation}
where  $k_{pl}$ is the Li plating kinetic rate constant, $c_{e}$ is the nominal concentration in electrolyte, and $c_{pl}$ is the concentration of plated lithium. Similar to the SEI case, we can simplify the Eqs.\ (\ref{plating system eqs}) into
\begin{flalign}
    \dot{\delta}_{pl}= \Omega_{pl} \: j_{pl} = f_{2}(\delta_{pl},C_n,LLI,I).
    \label{del pl dot}
\end{flalign}
Plated lithium moles can be calculated as
\begin{flalign}
    \Delta n_{Li, pl}= -\frac{A l^- a_{s}^- }{\Omega_{pl}}\delta_{pl}.
    \label{plated moles}
\end{flalign}

 \subsection{Mechanical degradation}

Intercalation of lithium into the electrodes causes them to expand, while deintercalation causes them to shrink. This repetitive expansion and shrinkage create oscillating stress within the electrodes, ultimately leading to cracking.
To model mechanical degradation, we use a crack growth model based on data-driven material fatigue models. The derivation of formulas for this stress is complex, and we only show the results of the derivations. For a more detailed description, we recommend referring to the work of \cite{pannala_movahedi_garrick_siegel_stefanopoulou_2023}.
The stresses in the particles ($\sigma_h = \sigma_h(c_{ss}, r_p$) can be assumed to be the function of the concentration profiles in each electrode, and the rate of loss of active material in each electrode can be presented as:
\begin{flalign}
        \dot{\varepsilon^\pm}= -\beta_1^\pm\left(\frac{|\sigma_{max}^\pm|}{\sigma_{crit}^\pm}\right)^{m_{LAM}} - \beta_2^\pm \left(\frac{|\sigma_{min}^\pm|}{\sigma_{crit}^\pm}\right)^{m_{LAM}}
        \label{eps dot}
\end{flalign}

where $\sigma_{max}=max(\sigma_h$) and$\sigma_{min}=min(\sigma_h$) in each cycle. Critical stress is shown by $\sigma_{critical}$, and $\beta_{1,2}$ and $m_{LAM}$ are coefficients related to the material characteristics of each electrode.
To simplify the notations, we can write Eq.\;(\ref{eps dot}) as: 
\begin{flalign}
\label{Cp dot}
    \dot{C_{p}}= f_{3} (C_{p}, I)\\
    \label{Cn dot}
    \dot{C_{n}}= f_{4} (C_{n}, I)
\end{flalign}

\subsection{Loss of lithium inventory}

The cyclable lithium inventory is calculated as
\begin{flalign}
    n_{Li}= \frac{3600}{F}(x C_n + y C_p)
    \label{nLi}
\end{flalign}
where $x$ and $y$ are the lithium stoichiometric at the negative and positive electrode. The LLI is equivalent to full cell capacity loss. However, framing the capacity loss as LLI enables us to further ascribe the mechanistic source of capacity loss since the lithium inventory is lost through multiple pathways as explained previously. Here, we develop analytical expressions for each pathway, including inventory loss due to side reactions and LAM. At cycle $k$, LLI can be calculated by dividing the loss in cyclable lithium by its initial amount:
\begin{gather}   
    LLI=\frac{\Delta n_{Li}}{n_{Li,0}}=\frac{n_{Li,0}-n_{Li,k}}{n_{Li,0}}
\end{gather}

 \cite{sulzer2021accelerated} have proven that it is possible to express the amount of lost Lithium moles as a combination of lithium lost due to side reactions (for example SEI growth and plating in this paper) and the lithium lost due to the LAM.
 
\begin{flalign}
     \Delta n_{Li}= \Delta n_{Li,SEI}+ \Delta n{Li, pl}+\frac{3600}{F}\int{(y \dot{C}_p+x \dot{C}_n)\ dt }
    \label{nLI due to side and LAM1}
\end{flalign}
where $\Delta n_{Li,SEI} \ \mathrm{and} \ \Delta n{Li, pl}$ are presented in Eqs. (\ref{moles SEI}, \ref{plated moles}) and we can define the lithium moles lost due to LAM as:
\begin{flalign}
    \Delta n_{Li, LAM}=\frac{3600}{F}\int{(y \dot{C}_p+x \dot{C}_n)\ dt }
\end{flalign}
If we look at the derivative of the LLI with respect to time:
\begin{flalign}
    \dot{LLI} = -A l^- a_{s}^- \left( 
    \frac{2\dot{\delta}_{SEI}}{\Omega_{SEI}} 
    +\frac{\dot {\delta}_{pl}}{\Omega_{pl}} \right)+ 
    \frac{3600}{F}{(y \dot{C}_p+x \dot{C}_n) }     
    \label{LLI dot}
\end{flalign}



Hence, the simplified version of Eq. (\ref{LLI dot}) can be presented as: 
\begin{flalign}
    L \dot{L}I= f_{5}(C_{p},C_{n},LLI, I)
\end{flalign}


\section{Dynamic system}

In this section, we present a state-space representation of the degradation dynamic system and identify the states of degradation which we are calling deepSOH.
\subsection{{deep}SOH}
Considering Eqs. (\ref{del dot sei f1}, \ref{del pl dot}, \ref{Cp dot}, \ref{Cp dot}, and \ref{LLI dot}), we see that the derivatives of the variables $\mathcal{X}=[\delta_{SEI}, \delta_{pl}, C_{p}, C_{n}, LLI]^T$ can be written as a function of these variables and the input current:
\begin{flalign} 
    \dot{\mathcal{X}}=
     \begin{bmatrix}
      \dot{\delta}_{SEI}\\
      \dot{\delta}_{pl}\\
      \dot{C_{p}}\\
      \dot{C_{n}}\\
      \dot{LLI}\\
     \end{bmatrix}
     =
     \begin{bmatrix}
      f_{1}(\delta_{SEI},C_n,LLI,I)\\
      f_{2}(\delta_{pl}, C_n,LLI,I)\\
      f_{3} (C_{p}, I)\\
      f_{4} (C_{n}, I)\\
      f_{5}(C_{p},C_{n},LLI, I)\\
     \end{bmatrix}
     = f(\mathcal{X}, I) \label{dynamic system}
\end{flalign}
The variables in $\mathcal{X}$ are linearly independent and, therefore, can be assumed to be the internal state variables of the degradation system. In other words, this set of variables is the minimum number of variables required to determine the state of the system and make predictions about its future. This representation of degradation mechanisms would not be possible if we were only relying on eSOH. 
Indeed, as evidenced in Fig. \ref{fig: deepSOH blocks}, eSOH can be assumed to be a subset of deepSOH. Therefore, determining the eSOH is necessary but might not be sufficient depending on the active degradation mechanism in the system.

It should be emphasized again that since deepSOH represents the states of the active degradation mechanism in the battery, these could be different from case to case. For example, if there had been transition-metal dissolution in the cathode we would have needed to add an additional state to deepSOH (for instance moles of dissolved metal). Conversely, if the cells were undergoing calendar aging solely through SEI growth, then a model order reduction would allow us to follow only one state, such as LLI or $\delta_{SEI}$, which would be equivalent in that case. 

\subsection{Measurements or Calculations}
After introducing the state vector and state equations, we will now consider some output measurements that could help us determine these states at a given point in time. Some possible measurements of the system include eSOH, instantaneous resistance, and irreversible expansion.

\subsubsection{eSOH} \label{eSOH subsub}
As was mentioned earlier, the electrode-specific SOH includes electrode capacities and stoichiometric limits. Generally, the eSOH is estimated based on pseudo-OCV (Low current input, similar to Fig. \ref{RPTs}) data, by solving the following equations:
\begin{flalign}
    V_{max} &= U^+(y_{100})-U^-(x_{100})\\
    V_{min} &= U^+(y_{0})-U^-(x_{0})\\
    C &= C_n\left(x_{100}-x_{0}\right) = C_p\left(y_{0}-y_{100}\right)
\end{flalign}
along with Eq. (\ref{nLi}).
\cite{lee2020electrode} also used dVdQ curves to improve the accuracy of the estimation.
Even though the eSOH are estimates, for the sake of simplicity, we consider them measurements here:
\begin{equation}
    y_\mathrm{eSOH} = [C_p, C_n, LLI]^T
    \label{y eSOH}
\end{equation}

\subsubsection{Resistance}
As the cell ages, its resistance increases. Instantaneous resistance can be simply measured by recording the voltage drop or jump across the cell in response to a step current.
The terminal voltage of the cell can be calculated as:
\begin{gather}
    V_T=U^+(c_{ss}^+)+\eta^+ - U^-(c_{ss}^-)+\eta^- +IR_{film} \label{terminal volatge}
\end{gather}

where $\eta$ is the intercalation overpotential and $R_{film}$ is the combination of the resistance of the SEI layer and plated lithium:
\begin{flalign}
    R_{film}= \frac{\delta_{SEI}}{\kappa_{SEI}}+\frac{\delta_{pl}}{\kappa_{pl}}
\end{flalign}
The ionic conductivities of the SEI layer and plated lithium are $\kappa_{SEI}$ and $\kappa_{pl}$, respectively. To find the $R_s$ we can take the derivative of terminal voltage ($R_s= \frac{dV_T}{dI}$) with regard to current, and since we are interested in the instantaneous drop in the voltage, we can ignore the OCP functions \citep{marquis2019asymptotic} and have: 
\begin{multline}
        R_s=R_{film }+ \frac{RT}{(1-\alpha)F}\\(\frac{\gamma^+}{\sqrt{(I\gamma^+ )^2+1}}+\frac{\gamma^-}{\sqrt{(I\gamma^- )^2+1}})\label{derv of butler}
\end{multline}
where 
$\gamma^\pm= \frac{r_p^\pm}{6(1-\alpha)l^\pm A i_0^\pm} \label{gamma}$.

The exchange current density  $i_0^\pm$ can be calculated as:
\begin{flalign}
i_{0}^\pm=k_0^\pm \left(\bar{c_e}\right)^{(1-\alpha)}\left(c_{s,max}^\pm-c_{ss}^\pm \right)^{(1-\alpha)}\left(c_{ss}^\pm\right)^\alpha.
\end{flalign}
where $k_0^\pm$ is the reaction rate.
Clearly, calculating the surface concentration requires solving the SPM equations. However, by assuming a small current step size, we can use the average concentration in each electrode ($  c_{ss}^\pm \approx c_{avg}^\pm = \frac{y/x}{c_{s, max}^\pm}$). 
Hence we can write the instantaneous resistance as:
\begin{gather}
    R_s=\frac{\delta_{SEI}}{\kappa_{SEI}}+\frac{\delta_{pl}}{\kappa_{pl}}+h_4(C_p,C_n)
    \label{Rs h2}
\end{gather}
\subsubsection{Irreversible expansion}
Lithium intercalation in electrode materials causes reversible volume changes in the crystal lattice, observable at the electrode and cell levels \citep{Pannala2022IFAC}. Additionally, irreversible thickness growth occurs due to SEI growth and lithium plating. Although lithium plating is reversible, some of the lithium can oxidize or disconnect, forming "dead-lithium" that contributes to irreversible thickness growth and increased transport losses in the electrolyte.


We have shown that the irreversible expansion can be modeled (\cite{pannala_movahedi_garrick_siegel_stefanopoulou_2023}) empirically using the thickness of the SEI layer, plating, and changes in the active material ratio. 
\begin{flalign}
    \label{irr expansion}
    \delta_{irrv}=b_{SEI}\delta_{SEI}+b_{pl}\delta_{pl}^2+ b_{in}^+LAM^+ + b_{in}^- LAM^-
\end{flalign}
where $b_{SEI}$, $b_{pl}$, and $b_{in}^\pm$ are scaling coefficients. To simplify the notation, we write Eq. (\ref{irr expansion}) as:
\begin{flalign}
        \label{irr expansion h3}
    \delta_{irrv}=b_{SEI}\delta_{SEI}+b_{pl}\delta_{pl}^2+ h_{5}(C_{p},C_{n})
\end{flalign}

 The cell irreversible expansion can be measured in the lab using a displacement sensor as in \cite{Mohtat2022, Pannala2022IFAC} or a load cell.
 In an automotive battery pack, the cells are often grouped in modules and constrained to a fixed volume within each module. To account for the reversible and irreversible expansion,  a compliant porous polyurethane pad is often placed adjacent to each pouch cell to absorb the cell dimensional changes.
 The increased thickness of the cell further compresses the pad (due to the constrained overall stack height) and increases the surface/normal pressure on the stack, which can be measured using a pressure sensor or load cell.


\section{Identifiability and simulations}

In this section, we aim to determine the states of the degradation system in (\ref{dynamic system}) using the output measurements obtained from cells of different ages, subjected to various aging protocols. We assume that we have no information about the aging process of each cell, but only know their initial condition when they were new.
\subsection{Measuring eSOH and resistance}
The first measurements we will consider are eSOH (Eq. \ref{y eSOH}) and resistance (Eq. \ref{Rs h2}).
\begin{gather}
    \label{Output Rs LLI}
    y=
     \begin{bmatrix}
     C_{P}\\
     C_{n}\\
     LLI\\
     R_s
     \end{bmatrix}
     = 
     \begin{bmatrix}
     C_{P}\\
     C_{n}\\
     LLI\\
     \frac{\delta_{SEI}}{\kappa_{SEI}}+\frac{\delta_{pl}}{\kappa_{pl}}+h_4(C_p,C_n)
     \end{bmatrix}
\end{gather}


It is clear that these outputs cannot uniquely determine the states of the system $\mathcal{X}=[\delta_{SEI}, \delta_{pl}, C_{p}, C_{n}, LLI]^T$
To be more specific, we cannot determine the states related to SEI and lithium plating. In other words, there are infinite sets of $[\delta_{SEI}, \delta_{pl}]$ that satisfy the measurement constraints in Eq.(\ref{Output Rs LLI}).

\begin{figure}[h]
\begin{center}
\includegraphics[width=6.7cm, height=16cm]{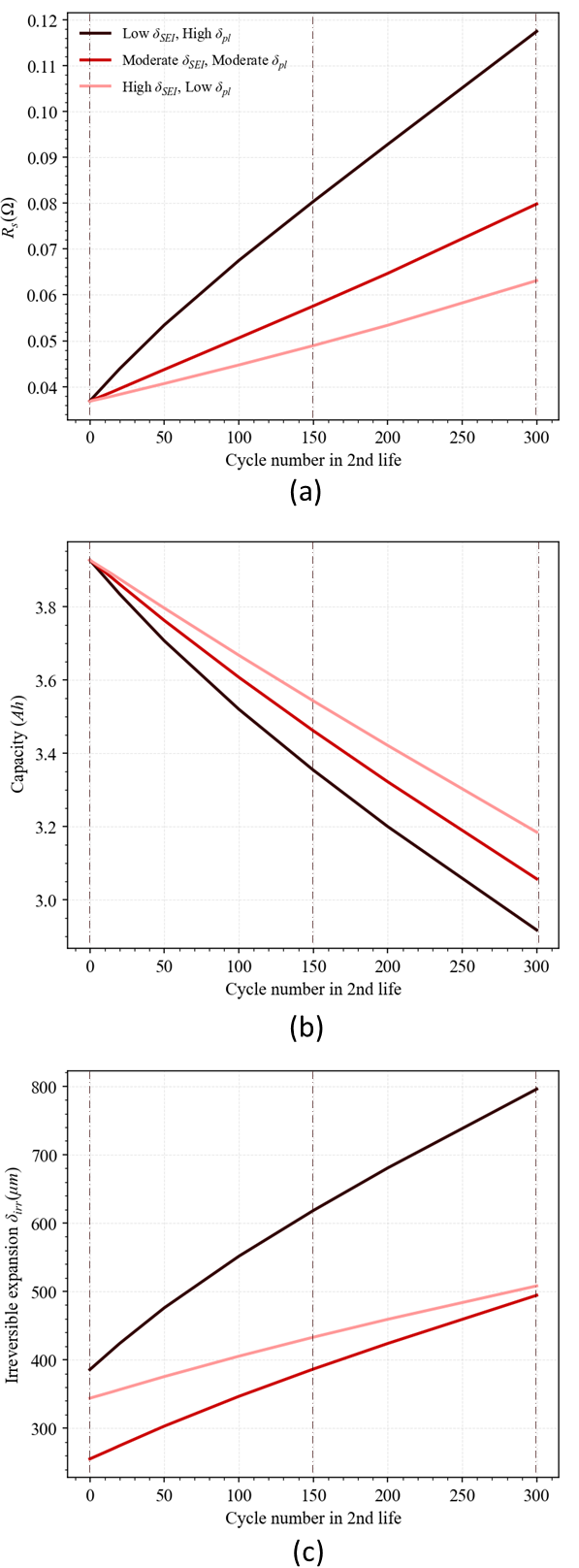}    
\caption{Possible degradation trajectories of cells with the same eSOH and resistance and different deepSOH values ($\delta_{SEI}$ and $\delta_{pl})$ in particular} 
\label{fig: ResCapExp}
\end{center}
\end{figure}
\begin{figure}[h]
\includegraphics[width=7cm, height=13cm]{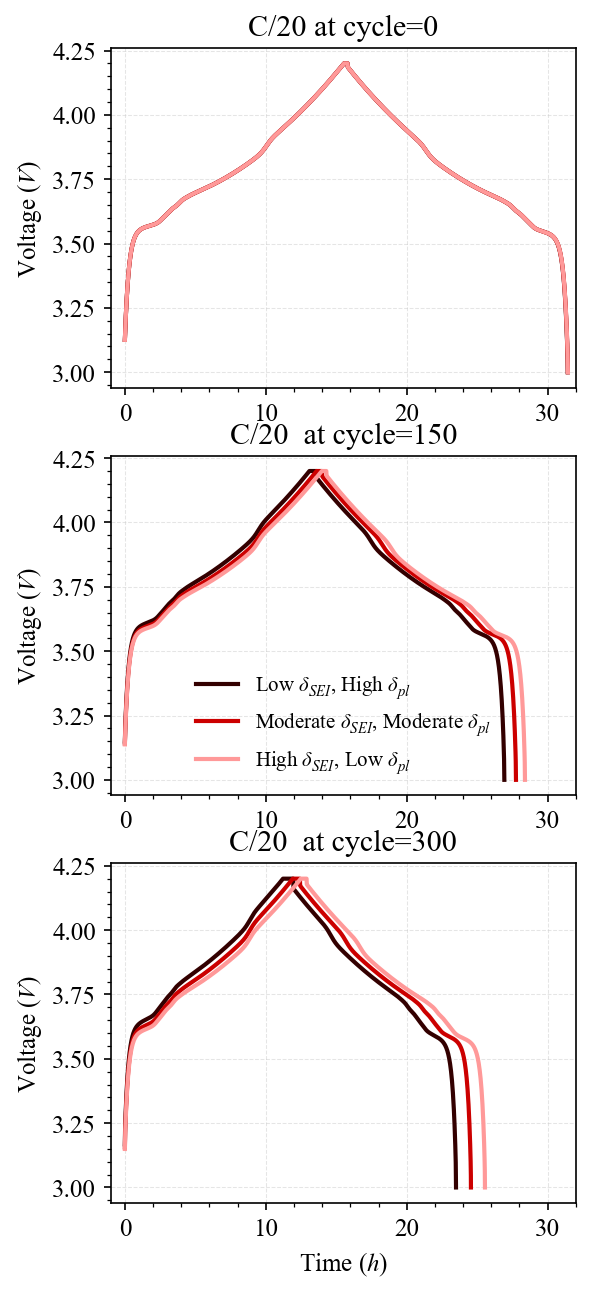}
  \caption{Voltage profile of C/20 charge and discharge of the cells the same initial eSOH and resistance and different deepSOH ($\delta_{SEI}$ and $\delta_{pl}$) values.}
  \label{RPTs}
\end{figure}

To further demonstrate this phenomenon, three aged cells with identical eSOH and instantaneous resistance values but different SEI and plated lithium thicknesses are modeled. The cells were simulated using PyBaMM \citep{sulzer2021python}. The second life cycling procedure used consisted of a discharge at C/5 until 3V, rest for 10s, charge at C/5 until 4.2V with a C/100 hold, and rest for 18hrs. The resulting resistance and capacity for these cells are shown in Fig. \ref{fig: ResCapExp} (a) and (b). As can be seen, even though these cells were identical in terms of resistance and eSOH at the same point in life, they have very different RULs. 
The voltage profiles of the cells at C/20 charge and discharge currents are shown in Fig. \ref{RPTs} at different stages of their second lives. This figure further demonstrates that the initial eSOH measurements for these three cells are identical. 
Since eSOH is estimated using the voltage output profiles of the cells during low-current charge/discharge and the voltage profiles of the three cells at the beginning (cycle=0) of their second lives are identical, their eSOH is the same.

These simulations, clearly show the insufficiency of the eSOH measurements for predicting the RUL. Even when resistance measurements are used in conjunction with eSOH measurements it cannot yield an accurate and unique description of the degradation system. 

Different trajectories of the cells in the second life can also be explained from a physical standpoint. It has been shown that Li-plating can accelerate trajectories \citep{yang2017modeling} and SEI growth can be self-limiting \citep{movahedi2023physics} trends. The nonlinearities present in the system will cause different paths for each of the cells. For instance, when considering possible sets of [$\delta_{SEI}$, $\delta_{pl}$], if we have a low $\delta_{SEI}$ and a high $\delta_{pl}$, then the SEI will increase significantly since the rate of self-limitation is still small. On the other hand, the plating rate is also higher. This particular combination will lead to a substantial increase in resistance and capacity. 

 
\subsection{Additional measurements}
What we demonstrated here, indicates the need for additional data for determining the deepSOH.
One approach to gathering this data is to consider additional output measurements. One possible additional measurement could be the irreversible expansion that the cells exhibit during aging. In other words combining Eq. (\ref{irr expansion}) and system of equations in (\ref{Output Rs LLI}) can result in: 

\begin{gather}
    \label{Output Rs LLI exp}
    y=
     \begin{bmatrix}
     C_{p}\\
     C_{n}\\
     LLI\\
     R_s\\
     \delta_{irr}
     \end{bmatrix}
     = 
     \begin{bmatrix}
     C_{p}\\
     C_{n}\\
     LLI\\
     \frac{\delta_{SEI}}{\kappa_{SEI}}+\frac{\delta_{pl}}{\kappa_{pl}}+h_4(C_p,C_n)\\
     b_{SEI}\delta_{SEI}+b_{pl}\delta_{pl}^2+ h_{5}(C_{p},C_{n})
     \end{bmatrix}
\end{gather}

The equations in (\ref{Output Rs LLI exp}) have either a unique solution or no solution. This fact is also evident from Fig.(\ref{fig: ResCapExp})(c).
This figure shows that although three cells have the same initial eSOH and resistance, they exhibit varying levels of irreversible expansion at the start of their second life.

An issue with relying on sparse measurements without heeding the dynamic of the system is that our estimate can be prone to noise and uncertainty in the dynamics (modeling error) and outputs (measurement noise or error stemming from the eSOH fitting method).  
 Another approach is to measure the system inputs (current load) to the system and estimate the states in real time. The observability of the system needs to be taken into account in that case.

\section{Conclusions}
In this paper, we introduced deepSOH as the state vector of the degradation mechanism states.
We showed the necessity for identifying these states to have a full picture of the degradation and uniquely determine the RUL.
To demonstrate this, we showed that cells with identical eSOH and even the same resistance values at a certain point in the battery's life can result in very different trajectories and RULs. 
Therefore, we should either add additional measurements such as irreversible expansion to increase the identifiability of deepSOH or record the use data of the battery of the cell to be able to determine the deepSOH.
Our analysis is based on leveraging a degradation mechanism model and assuming that we have complete knowledge of the cell dynamics and degradation mechanism parameters. However, in future works, we will address the parameterization and uncertainty in the cell and degradation models. 

\bibliography{main}

\begin{thebibliography}{14}
\providecommand{\natexlab}[1]{#1}
\providecommand{\url}[1]{\texttt{#1}}
\providecommand{\urlprefix}{URL }
\expandafter\ifx\csname urlstyle\endcsname\relax
  \providecommand{\doi}[1]{doi:\discretionary{}{}{}#1}\else
  \providecommand{\doi}{doi:\discretionary{}{}{}\begingroup \urlstyle{rm}\Url}\fi

\bibitem[{Attia et~al.(2022)Attia, Bills, Planella, Dechent, dos Reis, Dubarry, Gasper, Gilchrist, Greenbank, Howey, Liu, Khoo, Preger, Soni, Sripad, Stefanopoulou, and Sulzer}]{Attia2022-mf}
Attia, P.M., Bills, A., Planella, F.B., Dechent, P., dos Reis, G., Dubarry, M., Gasper, P., Gilchrist, R., Greenbank, S., Howey, D., Liu, O., Khoo, E., Preger, Y., Soni, A., Sripad, S., Stefanopoulou, A.G., and Sulzer, V. (2022).
\newblock {Review---``Knees''} in {Lithium-Ion} battery aging trajectories.
\newblock \emph{J. Electrochem. Soc.}, 169(6), 060517.

\bibitem[{Butts(2024)}]{Butts2024-ed}
Butts, D. (2024).
\newblock Poor resale values of {EVs} threaten adoption, warn some experts.
\newblock Accessed: 2024-4-22.

\bibitem[{Lee et~al.(2020)Lee, Siegel, Stefanopoulou, Lee, and Lee}]{lee2020electrode}
Lee, S., Siegel, J.B., Stefanopoulou, A.G., Lee, J.W., and Lee, T.K. (2020).
\newblock {Electrode state of health estimation for Lithium ion batteries considering half-cell potential change due to aging}.
\newblock \emph{Journal of The Electrochemical Society}, 167(9), 090531.

\bibitem[{Lopetegi et~al.(2024)Lopetegi, Plett, Trimboli, Oca, Miguel, and Iraola}]{lopetegi2024new}
Lopetegi, I., Plett, G.L., Trimboli, M.S., Oca, L., Miguel, E., and Iraola, U. (2024).
\newblock A new battery soc/soh/esoh estimation method using a pbm and interconnected spkfs: Part ii. soh and esoh estimation.
\newblock \emph{Journal of The Electrochemical Society}, 171(3), 030518.

\bibitem[{Marquis et~al.(2019)Marquis, Sulzer, Timms, Please, and Chapman}]{marquis2019asymptotic}
Marquis, S.G., Sulzer, V., Timms, R., Please, C.P., and Chapman, S.J. (2019).
\newblock An asymptotic derivation of a single particle model with electrolyte.
\newblock \emph{Journal of The Electrochemical Society}, 166(15), A3693.

\bibitem[{Mohtat et~al.(2019)Mohtat, Lee, Siegel, and Stefanopoulou}]{mohtat2019towards}
Mohtat, P., Lee, S., Siegel, J.B., and Stefanopoulou, A.G. (2019).
\newblock Towards better estimability of electrode-specific state of health: Decoding the cell expansion.
\newblock \emph{Journal of Power Sources}, 427, 101--111.

\bibitem[{Mohtat et~al.(2022)Mohtat, Lee, Siegel, and Stefanopoulou}]{Mohtat2022}
Mohtat, P., Lee, S., Siegel, J.B., and Stefanopoulou, A.G. (2022).
\newblock Comparison of expansion and voltage differential indicators for battery capacity fade.
\newblock \emph{Journal of Power Sources}, 518, 230714.
\newblock \doi{10.1016/j.jpowsour.2021.230714}.

\bibitem[{Mohtat et~al.(2017)Mohtat, Nezampasandarbabi, Mohan, Siegel, and Stefanopoulou}]{mohtat2017identifying}
Mohtat, P., Nezampasandarbabi, F., Mohan, S., Siegel, J.B., and Stefanopoulou, A.G. (2017).
\newblock On identifying the aging mechanisms in li-ion batteries using two points measurements.
\newblock In \emph{2017 American Control Conference (ACC)}, 98--103. IEEE.

\bibitem[{Movahedi et~al.(2023)Movahedi, Pannala, Siegel, and Stefanopoulou}]{movahedi2023physics}
Movahedi, H., Pannala, S., Siegel, J.B., and Stefanopoulou, A.G. (2023).
\newblock {Physics-informed optimal experiment design of calendar aging tests and sensitivity analysis for SEI parameters estimation in Lithium-ion batteries}.
\newblock \emph{IFAC-PapersOnLine}, 56(3), 433--438.
\newblock \doi{10.1016/j.ifacol.2023.12.062}.

\bibitem[{Pannala et~al.(2023)Pannala, Movahedi, Garrick, Stefanopoulou, and Siegel}]{pannala_movahedi_garrick_siegel_stefanopoulou_2023}
Pannala, S., Movahedi, H., Garrick, T.R., Stefanopoulou, A., and Siegel, J. (2023).
\newblock Consistently tuned battery lifetime predictive model of capacity loss, resistance increase, and irreversible thickness growth.
\newblock \emph{Journal of The Electrochemical Society}.
\newblock \doi{10.1149/1945-7111/ad1294}.

\bibitem[{Pannala et~al.(2022)Pannala, Weng, Fischer, Siegel, and Stefanopoulou}]{Pannala2022IFAC}
Pannala, S., Weng, A., Fischer, I., Siegel, J.B., and Stefanopoulou, A.G. (2022).
\newblock Low-cost inductive sensor and fixture kit for measuring battery cell thickness under constant pressure.
\newblock \emph{IFAC-PapersOnLine}, 55(37), 712--717.
\newblock \doi{10.1016/j.ifacol.2022.11.266}.

\bibitem[{Sulzer et~al.(2021{\natexlab{a}})Sulzer, Marquis, Timms, Robinson, and Chapman}]{sulzer2021python}
Sulzer, V., Marquis, S.G., Timms, R., Robinson, M., and Chapman, S.J. (2021{\natexlab{a}}).
\newblock {Python battery mathematical modelling (PyBaMM)}.
\newblock \emph{Journal of Open Research Software}, 9(1).

\bibitem[{Sulzer et~al.(2021{\natexlab{b}})Sulzer, Mohtat, Pannala, Siegel, and Stefanopoulou}]{sulzer2021accelerated}
Sulzer, V., Mohtat, P., Pannala, S., Siegel, J.B., and Stefanopoulou, A.G. (2021{\natexlab{b}}).
\newblock Accelerated battery lifetime simulations using adaptive inter-cycle extrapolation algorithm.
\newblock \emph{Journal of The Electrochemical Society}, 168(12), 120531.
\newblock \doi{10.1149/1945-7111/ac3e48}.

\bibitem[{Yang et~al.(2017)Yang, Leng, Zhang, Ge, and Wang}]{yang2017modeling}
Yang, X.G., Leng, Y., Zhang, G., Ge, S., and Wang, C.Y. (2017).
\newblock Modeling of lithium plating induced aging of lithium-ion batteries: Transition from linear to nonlinear aging.
\newblock \emph{Journal of Power Sources}, 360, 28--40.

\end{thebibliography}
\end{document}